\documentclass{ws-ijmpd}
\usepackage[super,compress]{cite}
\usepackage{hyperref}
\hypersetup{
  colorlinks=true,        
  linkcolor=blue,         
  citecolor=cyan,         
}

\usepackage{graphicx}
\begin{document}

%
\catchline{}{}{}{}{}
%

\title{On black hole formation in higher dimensions}

\author{NARESH DADHICH }

\address{Inter University Centre for Astronomy \&
Astrophysics, Post Bag 4, Pune 411007, India \\ nkd@iucaa.in}

\author{SANJAR SHAYMATOV }

\address{Institute for Theoretical Physics and Cosmology, Zheijiang University of Technology, Hangzhou 310023, China \\ Akfa University, Kichik Halqa Yuli Street 17, Tashkent 100095, Uzbekistan\\
Ulugh Beg Astronomical Institute, Astronomy St. 33,
    Tashkent 100052, Uzbekistan \\ National University of Uzbekistan, Tashkent 100174, Uzbekistan \\
Tashkent Institute of Irrigation and Agricultural Mechanization Engineers, Kori Niyoziy 39, Tashkent 100000, Uzbekistan\\ sanjar@astrin.uz}

\maketitle

\begin{history}
\received{Day Month Year}
\revised{Day Month Year}
\end{history}

\begin{abstract}
The two main processes of black hole formation are: one, collapse of a matter cloud under its own gravity and the other is accretion of matter onto an already existing gravitating centre. The necessary condition for both the processes to operate is that overall force on collapsing fluid element or on test accreting particles is attractive. It turns out that this is not the case in general in higher dimensions greater than the usual four for collapsing or accreting matter having non-zero angular momentum. Thus both these processes cannot operate in higher dimensions to form a rotating black hole. The only theory in which this is not the case in higher dimensions is the pure Lovelock gravity where both these processes could in principle work for formation of black holes.

\end{abstract}


\ccode{PACS numbers: }


\section{Introduction}
\label{introduction}

The two main processes responsible for formation of black holes are gravitational collapse and accretion onto a gravitating centre. For the former one has to dynamically evolve gravitational collapse of matter distribution with equation of state from given regular initial data. This is highly complex and involved process. While for the latter, one has the benefit of exterior metric for setting up accretion process. The process involves matter revolving around the centre in a disk which would always occur so long as there exist stable circular orbits (SCOs). In the Newtonian theory, they exist everywhere while for general relativity (GR) there exists the  minimum radius for the innermost stable circular orbit (ISCO). Its angular momentum defines the lower bound on particle's angular momentum to execute an SCO. SCO in accretion disk keeps on falling inwards as it loses angular momentum by dissipative viscous forces present there. That is how matter keeps on accreting on central object positing angular momentum on it.

The primary and necessary requirement for both these processes to work is that overall force involving both gravitational attraction and repulsive centrifugal components is attractive. If that is not the case, overall force on fluid element for collapse and for test particle for accretion would be repulsive and hence both these processes would not work or rather work in the reverse. In this letter we wish to demonstrate this explicitly in higher dimensions.

Accretion process is facilitated by accretion disk in which matter revolves around the gravitating centre in SCOs. Thus existence of SCOs becomes the critical condition for accretion to ensue. It is rather well known that in GR as well as in Newtonian gravity, no bound orbits and thereby SCOs can occur in higher dimensions greater than the usual four. The reason is very simple, gravitational potential in $D$-dimension goes as $1/r^{D-3}$ while the centrifugal potential as always falls off as $1/r^2$. In $D>4$, the former would not be able to counter balance the latter to produce a potential well so as to harbour bound orbits. That is, effective potential has no minimum and consequently no stable circular orbit.

In contrast to Newtonian gravity in GR rotation also contributes to gravitational potential as well as to the phenomenon of frame dragging. That is, rotation does not remain confined only to black hole but it is also shared by space around it, causing a zero angular momentum  particle to have non-zero angular velocity --- dragging of inertial frame. However both these effects die out sharply with increasing $r$. In higher dimensions a rotating black hole is described the Myers-Perry solution of vacuum equation \cite{Myers-Perry86}. It turns out that as for non-rotating black hole, bound orbits and thereby SCOs cannot exist for higher dimensional rotating black hole as well. This is because effective potential for large $r$ reduces to the same form as that due to non-rotating black hole.

Since there can occur no SCOs, there can be no accretion disk to facilitate accretion process. On the other hand for gravitational collapse is more involved process for which one has to study fully relativistic evolution of fluid cloud from a regular initial data. It has to have angular momentum so as to form a rotating black hole. The primary requirement for collapse to ensue is that the overall force on rotating fluid element must be attractive. Since collapse is to start from large $r$ where resultant force would be $l^2/{2r^2} - M/{2r^{D-3}}$ which would clearly be repulsive for $D>4$. Thus it would not satisfy the necessary condition for collapse to begin, and therefore there is no question of further study of its evolution.

Thus both gravitational collapse and accretion cannot work for formation of rotating black holes in higher dimensions. We would employ the Myers-Perry metric \cite{Myers-Perry86} to prove the above assertions. This is the principal aim of this letter.

 This would also have a bearing on the phenomenon of overspinning of black hole which was initiated by Ref.~\cite{Jacobson09}. Since then it  has been pursued quite intensely as evidenced by large number of papers (see some representative
examples~\cite{Barausse10,Bouhmadi-Lopez10,Saa11,Li13,Rocha14,
Colleoni15b,Shaymatov15}). It turns out that it is in general possible to overspin a black hole under linear test particle accretion\footnote{It is however not possible to convert a non-extremal into extremal black hole \cite{Dadhich97}, and nor an extremal into over-extremal state \cite{Wald74b} by geodetic accretion. What is envisaged here is that extremality cannot be reached but it could perhaps be jumped over by a discrete non-geodetic but linear order perturbative process.}. Recently in a breakthrough paper, Sorce and Wald \cite{Sorce-Wald17} have shown that this result is always overturned when non-linear perturbations are included. Thus weak cosmic censorship conjecture (WCCC) which may be violated at linear order is always restored at non-linear order.

The question arises, what happens in higher dimensions, how does the phenomenon of overspinning fare for the Myers-Perry rotating black holes? As pointed out earlier that overall force is repulsive in dimensions greater than five and hence it is expected that black hole may not be able to overspin even at the linear order. This is precisely what has been shown \cite{Shaymatov20a} with an explicit calculation. In all dimensions $D\geq6$ a black hole cannot be overspun in \cite{Shaymatov21a}. In five dimension it could however be overspun at linear order which gets overturned when second order perturbations are included \cite{Shaymatov19a,Shaymatov19c,Shaymatov21d}. Thus higher dimensions favour WCCC for rotating black holes.

We have also investigated \cite{Shaymatov20-pl} overspinning of pure Lovelock\footnote{Pure Lovelock means the Lovelock Lagrangian and the following equation of motion has only one $N$th order term without sum over lower orders. Here $N$ is degree of homogeneous Riemann curvature polynomial in action. It should however be noted that the metric we employ for rotating black hole is extrapolated \cite{Dadhich-Ghosh13} from the corresponding Myers-Perry metric but it is not an exact solution of pure Lovelock vacuum equation. It however satisfies the equation in the leading order and has all the desirable and expected features.} rotating black hole and have shown that it cannot be overspun by linear order accretion in dimension $>4N+1$. For $N=1$ Einstein gravity, it implies that a rotating black hole in $D>5$ cannot be overspun.

The paper is organized as follows: In Sec.~\ref{sec:metric} we briefly describe the higher dimensional Myers-Perry rotating black hole metric.  In Sec.~\ref{sec:eff} we build up the effective potential for test particles motion for $(D=5, 6)$-dimensional black holes. We end with discussion in the Sec.~\ref{sec:discussion}. Throughout we use a system of units in which $G=c=1$.

\section{Myers-Perry rotating black hole}\label{sec:metric}

The metric describing the higher dimensional Myers-Perry rotating black hole \cite{Myers-Perry86} is given by
\begin{eqnarray}\label{2n+1}
ds^2&=&-dt^2 + (r^2+a^2_{n})\left(d\mu_{n}^2+\mu_{n}^2d\phi^2_{n}\right)\nonumber\\&+&
\frac{M r^{2n+3-D}}{\Pi F}\left(dt +a_{n}\mu_{n}^2d\phi_{n}\right)^2 +\frac{\Pi F}{\Delta}dr^2 \nonumber\\&+& (D-2n-1)r^2 d\alpha^2 \, ,
\end{eqnarray}
where
\begin{eqnarray}
\Delta &=& \Pi - M r^{2n+3-D}\, , \nonumber \\
F &=& 1-\frac{a_{n}^2\mu_{n}^2} {r^2+a_{n}^2}\, , \nonumber\\
\Pi &=&\sum_{i=1}^{n}(r^2+a_i^2)\, ,
\end{eqnarray}
where $n= [(D-1)/2]$ is maximum number of rotation parameters a black hole can have in $D$ dimensions (see for example \cite{Myers-Perry86}) and $\Sigma \mu_n^2 + (D-2n-1)\alpha^2=1$. In odd $D=2n+1$ dimensions, $\Delta = \Pi- M r^2$ and $\Sigma \mu_n^2=1$ while for even $D=2n+2$, $\Delta = \Pi- M r$ and $\Sigma \mu_n^2 + \alpha^2=1$ where $M$ and $a$ are respectively mass and rotation parameters. The black hole horizon is given by positive root of $\Delta=0$.

It should however be noted that whenever one or more of rotation parameters are switched off, $\Delta = 0$ has only one positive root indicating occurrence of only one horizon~\cite{Shaymatov21a}. As argued earlier for our purpose it would suffice to consider only one  rotation parameter. In particular we would examine the case of $5$ and $6$ dimensions and show that (i) effective potential, $V_{eff} >1$ always for non-zero angular momentum, and (ii) it has only a maximum and no minimum and hence there can occur no bound and stable circular orbits. Following the standard procedure for geodesic motion of timelike particles in the equatorial plane around a rotating black hole, we would write the effective potential.

Horizon for five dimension is given by $r_+ = \sqrt{M - a^2}$ while for six dimension it reads as
\begin{eqnarray}\label{Eq:6d_single}
r_{+}&=& \left(\frac{M A}{2}\right)^{1/3}\left[1- \frac{1}{3}\left(\frac{2}{A}\right)^{2/3}\left(\frac{a}{M^{1/3}}\right)^2\right]\, ,
\end{eqnarray}
where
\begin{eqnarray}\label{Eq:6d_single}
 A &=& 1 + \sqrt{1 + \frac{4}{27} \left(\frac{a^3}{M}\right)^2} \, .
\end{eqnarray}

\section{Effective potential}\label{sec:eff}

\begin{figure*}
\centering

 \includegraphics[width=0.45\textwidth]{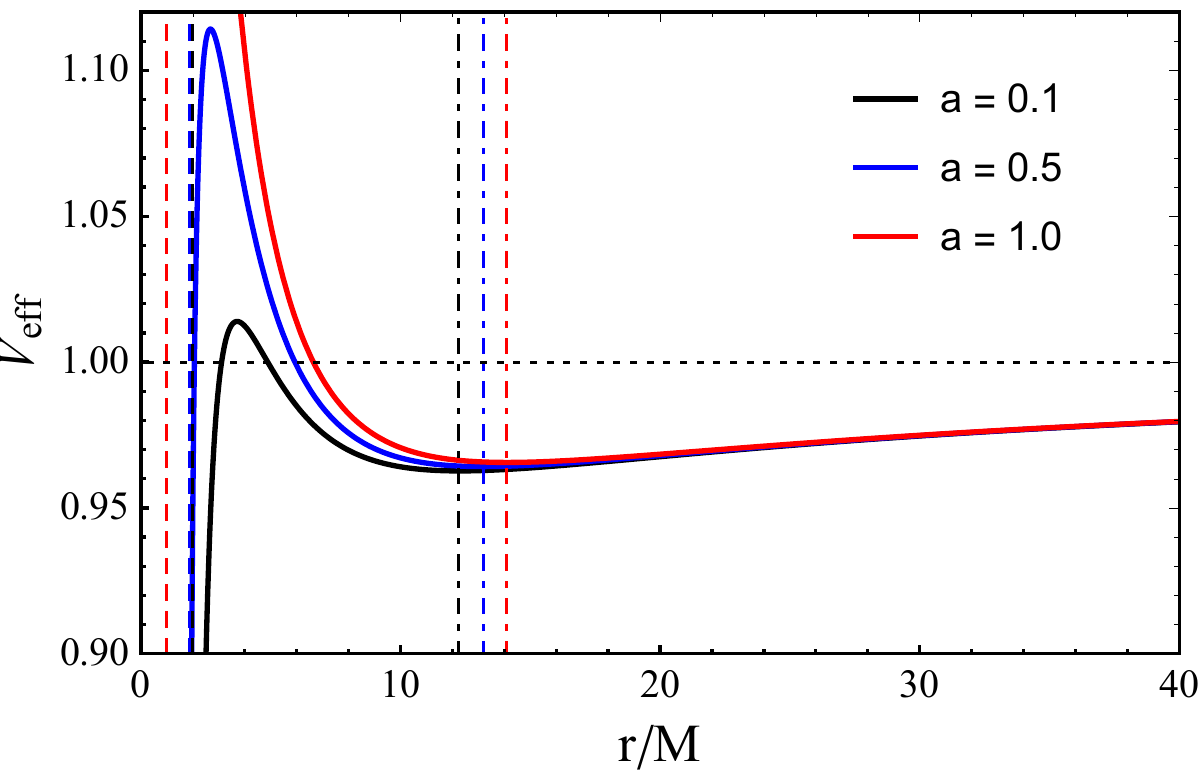}
 \includegraphics[width=0.45\textwidth]{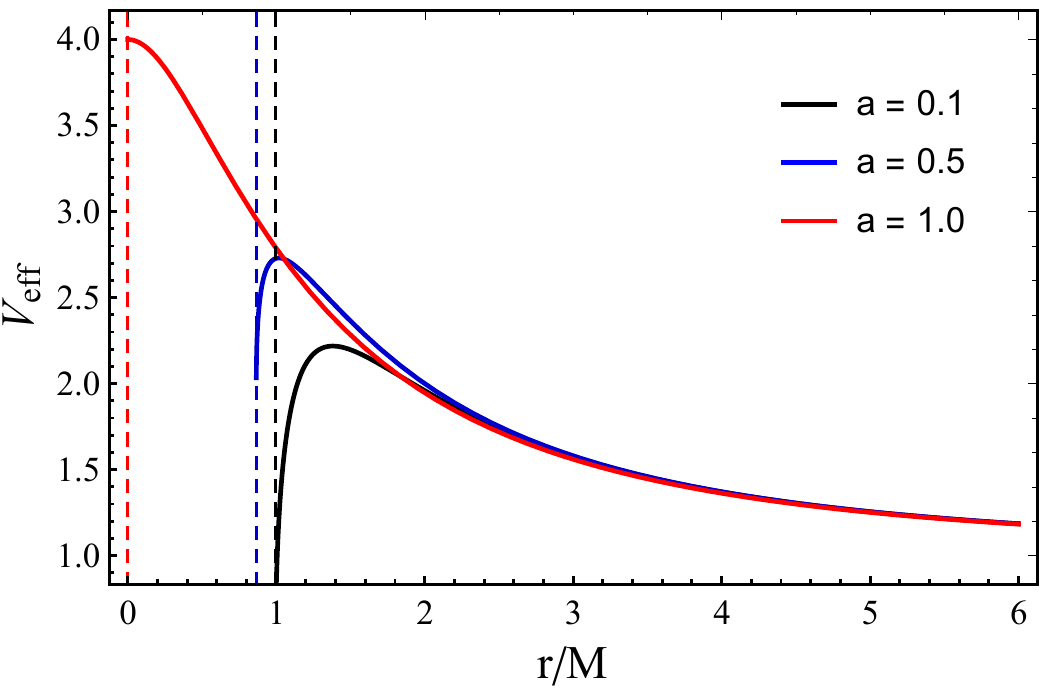}
 \includegraphics[width=0.45\textwidth]{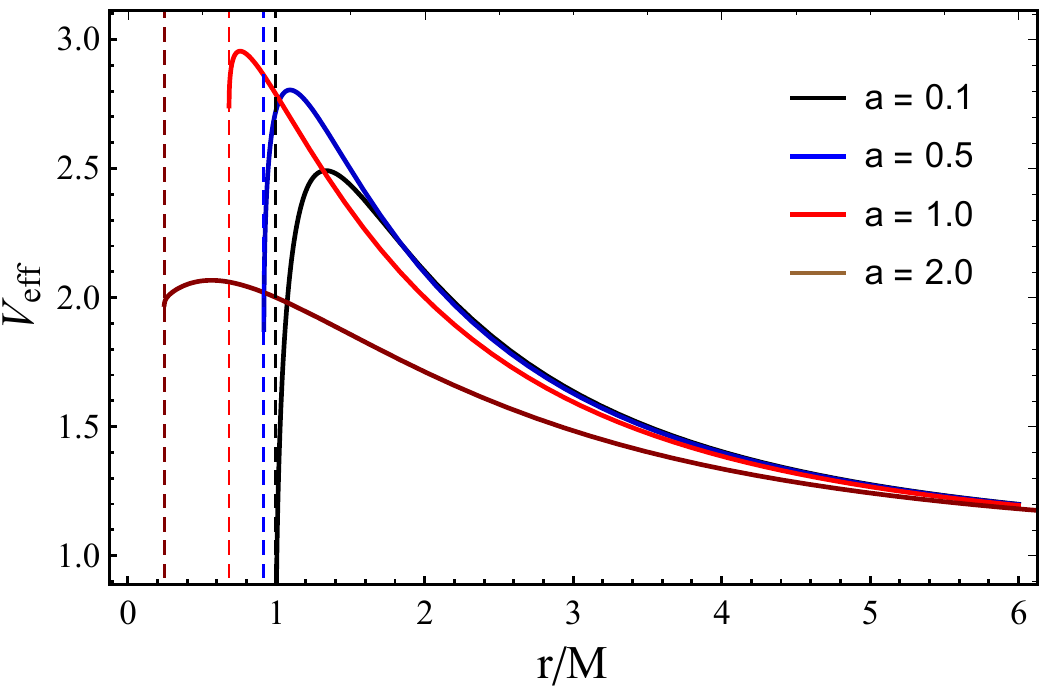}
\caption{\label{fig1} Effective potential plots for $L=4$: Left, right and below panels respectively refer to $D= 4, 5, 6$. The vertical dashed lines indicate location of horizon while the vertical dot-dashed lines indicate location of minimum of $V_{eff}$.}
\end{figure*}
The effective potential for timelike radial motion in equatorial plane for a rotating black hole with a single rotation is then generically given by
\begin{eqnarray}\label{Eq:6d_single}
V_{eff} = \Omega L + \sqrt{\frac{\Delta}{g_{\phi\phi}r^{2(n-1)}\, }\Big(L^2/g_{\phi\phi} + 1\Big)} \, ,
\end{eqnarray}
where $L$ is angular momentum of particle and  $\Omega = -g_{t\phi}/g_{\phi\phi}$ is the frame dragging angular velocity. The above expression follows from the geodesic equation of the Kerr geometry \cite{Wagh89} which is generalized to higher dimensions. This would explicitly read as follows:

\begin{eqnarray}
V_{eff}&=& \Phi\beta \frac{aL}{r^2}
+ \frac{\sqrt{\Big(\beta + \frac{L^2}{r^2}\Big)\Big(1 + \frac{a^2}{r^2} - \Phi\Big)}}{\beta} \, ,
\end{eqnarray}
where
\begin{eqnarray}
\Phi = \frac{M}{r^{D-3}}\, \mbox{~and~} \beta = 1 + \frac{a^2}{r^2} + \Phi \frac{a^2}{r^2} \, .
\end{eqnarray}

On expanding it for large $r$, it reduces to 
\begin{eqnarray}
V_{eff}&=& 1 + L^2/{2r^2} - M/{2r^{D-3}}\, ,
\end{eqnarray}
which clearly shows that the repulsive centrifugal component would override the attractive gravitational one for $D > 4$. Note that when $L=0$, effect of black hole rotation dies out sharply leaving only the attractive component. That is why all curves in potential plots in Fig.~\ref{fig1} merge for large enough $r$. 

For both $D = 5, 6$ $V_{eff}=1$ at infinity, and then it rises as $r$ decreases and reaches maximum before coming down at horizon. It is interesting  that $V_{eff}\geq 1$ all through except very close to horizon. This is in contrast to the four-dimensional case where $V_eff \leq 1$ away from horizon. For a single rotation, there occurs only one horizon and hence there is no upper limit on rotation parameter $a$ except for five dimension where it has to respect $a^2\leq M$ lest it turns into a naked singularity \footnote{This happens only in the special case of five dimension and not in general for $D=2n+1$, because in this case contribution to potential due to both mass and rotation falls as $1/r^2$.}. This is an interesting case of a rotating black hole with one horizon yet having an extremal limit for its rotation parameter. It is also interesting to note that in $D=6$, as $a\to\infty$, $V_{eff} \to 1$ at all $r$. This is why initially maximum of curve rises with increasing $a$ until $a\sim 1.3$, then it starts coming down.

\section{Discussion}\label{sec:discussion}

In Fig.~\ref{fig1} above, we have plots of effective potential for $L = 4$ in $D=4, 5, 6$. For the zero angular momentum case, the potential for large $r$ goes as $1 - M/2r^{D-3}$ and hence it would always be $\leq1$ (specific energy of particle at infinity is set to unity). That is, asymptotically contribution due to black hole rotation fades out, leaving only the one due to mass. What distinguishes four dimension (upper left panel) from $D>4$ (upper right and lower panels) is the fact that $V_{eff} \leq 1$ while in contrast it is opposite for the others. As a matter of fact it is greater than unity everywhere except near the horizon in $D>4$, reaching unity from the above and has only a maximum and no minimum. That means there can exist no potential well to harbour bound and SCOs. In contrast for the four dimension, since $V_{eff}$ reaches unity from the below, it has a minimum giving rise to a potential well for harbouring bound and stable circular orbits. This result was however known for non rotating black holes in higher dimensions \cite{Dadhich13}. It is due to the fact that centrifugal repulsive component overrides attractive one due to mass and thereby not letting the potential to have a minimum to form a well for bound orbits. This feature gets carried forward to rotating black holes because for large $r$, $V_{eff} \to 1 +L^2/{2r^2} - M/{2r^{D-3}}$ is the same as for non-rotating black hole. Thus it is no surprise that bound orbits and thereby SCOs cannot exist around rotating black holes in higher dimensions.

This raises the critical question about accretion process in higher dimensions. Accretion is mediated through accretion disk which cannot occur because there exist no bound orbits and consequently no SCOs. This is true for both rotating as well as non-rotating black holes in higher dimensions. Accretion disk provides avenue for dissipative interactions involving viscosity and collisions between particles through which particles can lose angular momentum and keep on falling inward and spiral into the hole with $L<L_{ISCO}$\footnote{Note that $L_{ISCO}$ defines the minimum threshold for particle to ride on a stable circular orbit. Hence particles with $L < L_{ISCO}$ will fall into black hole carrying angular momentum.}.  Since stable circular orbits cannot exist in higher dimensions for accretion disk to form, hence an accretion process cannot ensue. It can therefore play no role in formation of a rotating black hole in higher dimensions.

This however does not rule out a possibility of particle with angular momentum and energy exceeding the threshold determined by the maximum of potential barrier falling into black hole as studied in Refs.~\cite{Bouhmadi-Lopez10,Shaymatov21a} for overspinning of higher dimensional black hole. It was shown that overspinning was not possible because particles with overspinning parameters were not able to reach the horizon. Since $V_{eff} >1$ always, hence they would not be able to reach rotating black hole horizon unless they were somehow energised to a value overriding maximum of the potential barrier. The only possibility of such an energising process could perhaps be collision with other compact objects, like neutron stars or black holes. It would be a very complex and involved process which would require detailed simulation of collision process.

How about taking the question to generalized theories of gravity? The most natural generalization of general relativity in higher dimensions is the Lovelock theory which is quintessentially higher dimensional. It is only the pure Lovelock theory \cite{Dadhich16a}, which has only one $N$th order term without sum over lower orders in the Lovelock Lagrangian, that admits bound orbits in the dimension range $2N+1<D<4N+1$ \cite{Dadhich13} where $N$ is degree of the Lovelock polynomial. For $N=1$, Einstein gravity bound orbits exist only in four dimension while for $N=2$ pure Gauss-Bonnet (GB)  gravity, they do for $D=6,7,8$. On the other hand non rotating black hole \cite{Dadhich13GB} is stable \cite{Gannouji19} only in dimensions $D\geq 3N+1$; i.e. for pure Gauss-Bonnet in seven and eight dimensions. That is, in these dimensions there would occur SCOs and hence accretion disk could exist and thereby accretion  process could play a role in formation of rotating black holes \footnote{In a separate paper we would be studying particle motion for pure GB rotating black hole and in particular obtain the threshold value of angular momentum given by ISCO \cite{Dadhich21}}. Like Kerr black hole in four dimension, rotating black holes in pure GB gravity could be formed by the usual accretion process with an accretion disk in $D= 7, 8$ (leaving out $D=6$ for which black hole would be unstable.). In general for the dimension window, $3N+1 \leq D \leq 4N$, the usual accretion process could therefore work for  formation of pure Lovelock rotating black holes. It is however another matter that there does not yet exist an exact solution for pure Gauss-Bonnet vacuum equation describing a rotating black hole \cite{Dadhich-Ghosh13}.

Let us now return to the question of gravitational collapse which we had alluded in the beginning. It involves a matter configuration with rotation collapsing under its own gravity from a regular initial data. Here spacetime geometry would not be fixed but would be dynamically evolving. This is a very complicated and involved problem requiring fully relativistic hydrodynamic evolution which has to be tackled numerically by sophisticated simulations. All this would be relevant only when the necessary condition for gravitational collapse to ensue is satisfied. That is that overall force, involving repulsive centrifugal and attractive gravitational components, must be attractive. Since $V_{eff} > 1$ in $D>4$, gravitational collapse cannot ensue and hence it cannot participate in rotating black hole formation process.

{There is an interesting limiting case considered in Ref.~\cite{Igata15} in which it is shown that six dimensional Myers-Perry rotating black hole with single rotation greater than the critical value $a/\mu^{1/3} \sim 1.628$ flattens out like a pancake such that its geometry near the rotational axis is locally a direct product of the four dimensional Schwarzschild geometry with two dimensional flat space. Since the geometry is locally four dimensional it would very well admit bound orbits and so also stable circular orbits \footnote{{Stable circular orbits exist not only for timelike particles but also for photons. This is because momentum in flat directions is non-zero which provides an effective non-zero mass.}} as is the case for four dimensional Schwarzschild black hole. Since SCOs could exist in this flattened out pancake, an accretion process could in principle set in. Would that not be a counter example to what we have so far discussed? }

{First and foremost, the question is, where from  comes an over critically spinning black hole so as to make geometry locally a product, $M^{4} \times R^{2}$? Essentially locally spacetime becomes four dimensional and it is no longer a generic six dimensional black hole spacetime. Locally it is no longer a six dimensional rotating black hole instead it now reduces to the usual four dimensional Schwarzschild black hole. In special constructions like this including black rings and multi black holes \cite{Emparan02PRL,Elvang04PRL,Emparan04,Emparan06,Kastor93,
Yumoto12,Ozdemir04}, bound orbits can indeed exist in higher dimensions. Our concern however is of generic black hole spacetimes in higher dimensions. As we have argued and shown that $V_{eff}\geq1$ in $D\geq6$, and it is this property which makes both collapse and accretion non-operational. Notwithstanding the special circumstance of pancaked black hole \cite{Igata15} which requires an over critically spinning black hole, a generic rotating black hole in dimensions greater than and equal to six cannot be formed by gravitational collapse and accretion in general relativity. } 

This leaves then the only possibility of collisions and mergers of massive objects like black holes and neutron stars. This process has also to counteract overall repulsive force. This could happen only in very special circumstances where colliding objects have very large momentum to overcome repulsive barrier. Such events could would be few and far between.

Thus for Einstein gravity the usual processes of gravitational collapse and accretion for formation of rotating black holes cannot work in higher dimensions. However they could indeed work in pure Lovelock gravity. In particular rotating black holes could in principle be formed in pure GB gravity in $D=7,8$ by gravitational accretion and collapse. Like kinematicity of gravity in all critical odd dimensions $D=2N+1$ and existence of bound orbits \cite{Dadhich16a}, the formation of rotating black holes in higher dimensions is yet another distinguishing feature that singles out pure Lovelock gravity.

\section*{Acknowledgments}
{We would like to thank the anonymous referee for pointing out Ref.~\cite{Igata15}.} ND wishes to thank Ajit Kembhavi for some useful discussions. SS acknowledges the support of Uzbekistan Ministry for Innovative Development.

\bibliographystyle{ws-ijmpd}
\bibliography{gravreferences}

\end{document}